\documentstyle[preprint,aps]{revtex}

\begin{document}

\draft

\preprint{June, 2003}

\title{Magnetic Alignment Process : A New Mechanism for \\
       Extracting Energy from Rotating Black Holes}

\author{Hongsu Kim\footnote{e-mail : hongsu@astro.snu.ac.kr}}

\address{Astronomy Program, SEES, Seoul National University, Seoul, 151-742, KOREA} 

\author{Hyun Kyu Lee\footnote{e-mail : hklee@hepth.hanyang.ac.kr} and
        Chul H. Lee\footnote{e-mail : chlee@hepth.hanyang.ac.kr}} 

\address{Department of Physics, Hanyang University, Seoul, 133-791, KOREA}

\maketitle

\begin{abstract}
As a complementary or companion process to the Blandford-Znajek mechanism for the rotational energy extraction from a
Kerr black hole to serve as a viable model for the central engine of quasar, AGN, and even GRB, the magnetic alignment process
is proposed. In contrast to the environment assumed in the Blandford-Znajek mechanism in which the rotating hole's spin
axis and the asymptotic direction of the magnetic field are aligned, this new process operates when they are generally
misaligned which obviously is the more natural situation likely to happen. The time scales over which these two processes
take place and the total radiated powers released at the horizon by them have been estimated and found to be nearly
comparable. This may imply that in a random, non-axisymmetric configuration, the magnetic alignment process would operate 
first during which some of the hole's rotational energy is released and then upon completion of the alignment, the 
Blandford-Znajek mechanism may follow and the rest of the hole's rotational energy, if any left, would be further
released.

\end{abstract}


\narrowtext

\begin{center}
{\rm\bf I. Introduction}
\end{center}

At present, one of the most probable candidates for the central engine of quasar, active galactic nuclei (AGN) and even gamma ray
bursts (GRB) is believed to be the gigantic rotating black hole-accretion disk system. The reliable practical mechanisms for the
extraction of rotational energy from the Kerr hole, however, is not a well-established issue yet (see [1] and references therein). 
In the present work, we would like to propose a new such mechanism that has not been realized in a serious manner yet. 
Consider the general case when the asymptotic direction of the uniform magnetic field and a rotating (Kerr) hole's spin direction are 
``misaligned''. Then according to the work by King and Lasota [2], which was motivated by the earlier work by Press [3], the asymptotically
uniform magnetic field $B$ exerts torque on the rotating Kerr hole that brings the hole's angular momentum $J$ into an eventual
alignment with the field. Where then does this ``oblique'' magnetic field come from ? We will come to this issue later on when we comment on
the interesting related work by Znajek [4]. Indeed, the principle that underlies this adjustment process can be easily understood as 
follows : obviously, the total angular momentum of the whole system consisting of just the rotating hole and the surrounding, uniform  
magnetic field should be conserved. This, then, implies that the associated net torque should vanish. As a result, what happens is that,
on the one hand, the uniform magnetic field exerts torque on the rotating hole and on the other, the hole also exerts equal and opposite
torque on the magnetic field as a whole such that the asymptotic direction of the field and the hole's spin direction together come
to an eventual alignment. The torque involved in this alignment process, however, is just the component perpendicular to the hole's
spin axis. Particularly, the component of the torque along the spin axis does more interesting job while the alignment process is
under way. In fact, this component of the torque parallel to the hole's spin axis plays the central role in our new mechanism (dubbed
the ``magnetic alignment'' process) for the rotational energy extraction from Kerr black holes that we are about to propose in this
work. We shall come back to this issue in some detail shortly. 

\begin{center}
{\rm\bf II. Magnetic alignment process versus Blandford-Znajek mechanism}
\end{center}

Speaking of the possible mechanisms for the rotational energy extraction
from Kerr black holes, there is a well-established proposal, known as the ``Blandford-Znajek mchanism'' [1]. Therefore, it would be
natural to begin our discussion by providing a brief description that distinguishes our magnetic alignment process from the 
conventional Blandford-Znajek process. Evidently, since the fundamental distinction would lie in the associated operational natures of
the two processes, it seems relevant that we first should give a precise description of the mechanism for the angular momentum and 
hence the rotational energy transfer from the rotating hole to the magnetic field in this magnetic alignment process and then compare it
with its counterpart in the Blandford-Znajek process. In the magnetic alignment process, this mechanism is strictly dictated by the
elementary mechanical principle. To see how it works, again concentrate on the component of the torque along the hole's spin axis $G^{z}$,
which is the relevant physical quantity in our discussion here. As discussed above, then, due to the total angular momentum conservation
(or equivalently, the vanishing net torque) of the rotating hole-magnetic field system, the hole also exerts the torque $(-G^{z})$ on the
magnetic field as a whole which is equal and opposite to that the magnetic field exerts on the hole. Thus the torque that acts against the
hole's angular momentum direction and hence brakes it and spins it down exactly cancels with the torque that spins up the swirl of the
magnetic field lines. As a result, in a naive sense, the magnetic field as a whole gains (extracts) just the same amounts of angular
momentum or rotational energy as that lost by the hole. Of course, in practice, it would be that only part of the rotational energy is
transferred to the magnetic field. The rest of the rotational energy lost by the hole but not transferred to the magnetic field would
be absorbed by the hole and deposited as a form of increase in the hole's ``irreducible mass''[8].
In contrast, the way how some of the hole's rotational energy is carried off by the magnetic field lines threading the hole in the
conventional Blandford-Znajek process is rather different. There the angular momentum and the rotational energy transport is
achieved in terms of the conservation of ``electromagnetic'' angular momentum and energy flux flowing from the hole to its
nearby field particularly when the so-called ``force-free condition''  is satisfied in the surrounding magnetosphere. It is
also amusing to note that the Blandford-Znajek mechanism can be translated into a simple-minded circuit analysis. In this
alternative picture, the current, flowing from near the pole toward the equator of the rotating black hole geometry and the
poloidal magnetic field together generate a ``magnetic braking'' torque which is antiparallel to the hole's spin direction. As a
result, the rotating hole, which is a part of the circuit, spins down and its angular momentum and the rotational energy are
extracted which is reminiscent of the Goldreich-Julian unipolar inductor model for a pulsar [12]. Indeed, the Blandford-Znajek 
model of the black hole magnetosphere was constructed in close analogy with the Goldreich-Julian model of the pulsar
electrodynamics.
For more details, we refer the reader to the original work of Blandford and Znajek [1] and some comprehensive review
articles [9] on this topic including the ``membrane paradigm'' approach [13]. As such, in order for the Blandford-Znajek process
to work, the local magnetic field lines should be firmly rooted on the horizon of the rotating hole (see the second reference of [1]) 
to carry away some of the hole's
rotational energy by means of this ``contact interaction''. Thus it is interesting to note that the restrictions such as the
force-free condition and the local contact of the electromagnetic field lines with the horizon are obviously not the prerequisite in
our magnetic alignment process which only concerns global, mechanical angular momentum conservation of the rotating
hole-magnetic field system. Nevertheless, it seems to be likely and hence natural to expect that part of the hole's angular
momentum or equivalently the rotational energy gained by the magnetic field in the magnetic alignment process would be released
eventually to infinitely far region essentially as the form of charged particle emission much like in the Blandford-Znajek
process. Therefore even in this magnetic alignment process, it might seem relevant to employ particularly the force-free
condition in practice. This is particularly so when the magnetic field is strong enough to render the vacuum break down to produce
electron-positron pairs in which case both the Blandford-Znajek and the magnetic alignment processes can serve as viable models to
account for the observed spectrum of gamma ray bursters (GRB). We will come back to this issue in some more detail later on. This is how the two
processes can be distinguished from operational side. Evidently, we also should mention that the actual distinction comes from the
initial relative orientation between the asymptotic direction of the magnetic field and the hole's spin direction. Were they
aligned with each other, this type of process would not take place to begin with and solely the Blandford-Znajek process alone may
operate. If on the other hand they were misaligned to some extent, 
then almost immediately this magnetic alignment process would operate first during which some of the hole's rotational
energy is released and then upon completion of alignment, the Blandford-Znajek process would follow and the rest of the hole's
rotational energy, if any left, would be further released. Indeed, the proposal of this picture, ``the magnetic alignment process
precedes and then the Blandford-Znajek process follows'', is based on the fact that indeed the Blandford-Znajek mechanism assumes
the environment in which the rotating hole's spin axis and the asymptotic direction of the magnetic field are aligned to produce a stationary
axisymmetric black hole magnetosphere. Nevertheless, we do not exclude the possibility that in reality, this magnetic alignment process 
and the Blandford-Znajek process may well be concurrent while the alignment is under way. 
Before we proceed with the details of the magnetic alignment mechanism, it seems relevant to comment
on the likelihood of our initial condition, i.e., the misalignment between the direction of magnetic
field and the spin axis of the hole. Indeed, if it is the accretion disk which essentially generates
and keeps the magnetic fields, then this oblique initial configuration may seem unlikely to happen.
Namely, any such initial misalignment would be removed by the well-known Bardeen-Petterson effect [14]
in an efficient manner. To be a little more concrete, the viscosity between differentially rotating
layers of the disk would generates a viscous torque and it decomposes into three parts ;
the spin-down torque reducing the angular momentum of the fluid element, the alignment torque
bringing the angular momentum of the fluid element into an eventual alignment with the hole's spin
and lastly the precession torque generating the precession of the fluid angular momentum about
the spin axis of the hole (much like the Lense-Thirring effect does). And particularly due to the
existence of the alignment torque, any possible initial misalignment of the inner part of the disk
and hence the direction of the magnetic field near the hole with the hole's spin axis will disappear
on a viscous time scale. Even for this case, however, it is interesting to note that provided the
viscous time scale is not much shorter than the GRB time scale, the magnetic alignment process still
have a good chance to account for GRB events since both the Bardeen-Petterson and the magnetic 
alignment processes would be concurrent then. Indeed, there are two species of GRBs known ; the
long duration bursts which last for about few hundreds of seconds and the short duration bursts
that last for about ten miliseconds. And thus the magnetic alignment process certainly has a good
chance to account for the short duration bursts even if we take into account the Bardeen-Petterson
effect. As we mentioned earlier, however, there could be yet different situations in which the
initial misalignment (that essentially motivated our present study) comes in a more natural sense.
And that is the argument due to Znajek [4] and it goes as follows. For example, the gravitational 
collapse of a star that has formed a black hole could leave behind a cloud of
plasma orbiting round the hole carrying any large-scale non-axisymmetric magnetic field embedded 
in it. Alternatively, such a field could be generated by a neutron star near the hole. 
And there is no reason why accretion should necessarily take place. Thus if these possibilities can
be taken seriously, our initial condition, the oblique configuration, may be regarded as being quite 
likely.     

\begin{center}
{\rm\bf III. The magnetic alignment process - Macroscopic version}
\end{center}

We now turn to the description of the quantitative aspect of this magnetic alignment process. In standard gravity text books 
(see, for instance, [8]), one can find that the torque exerted on a gravitating body (like black holes, neutron stars, etc.) 
by a field with stress tensor $T^{\mu\nu}$ in an asymptotically-flat spacetime is given by
\begin{eqnarray}
G^{l} = - \int_{S_{\infty}}dS_{j} \epsilon^{lik}T^{ji}x^{k}.
\end{eqnarray}
where $dS_{j}$ denotes the surface element of 2-sphere in the asymptotically-flat region. For the case at hand in which the
gravitating body is a rotating Kerr black hole and the field in the interbody region is an asymptotically uniform magnetic field
$B$, this torque has been computed as [2,4]
\begin{eqnarray}
G^{x} &=& - J\tau^{-1}\sin \gamma \cos \gamma, ~~~G^{y} = 0,
\nonumber \\ G^{z} &=& - J\tau^{-1}\sin^{2} \gamma
\end{eqnarray}
or in a non-relativistic vector notation, 
$\vec{G} = d\vec{J}/dt = (2/3)M(\vec{J}\times \vec{B})\times \vec{B}$ and
where $\tau = (3/2)B^{-2}M^{-1}$ is the alignment time scale. $M$, $J=Ma$ are the 
Arnowitt-Deser-Misner (ADM) mass and the angular momentum of Kerr black
hole and $\gamma $ denotes the angle between the asymptotic direction of the magnetic field and the hole's spin direction.
Since in eq.(2) the hole's spin direction is chosen to be along the $z$-axis, the component of the torque perpendicular to the
hole's spin axis, $G^{x}$, is associated with the alignment process while the component along the spin axis, $G^{z}$, has to
do with the angular momentum/rotational energy extraction process. Particularly note that the minus signs in eq.(2) indicate that
$G^{x}$ is an ``alignment'' torque rather than the other way around and $G^{z}$ is a ``braking'' torque rather than an accelerating one.
Lastly, the rationale for the vanishing $G^{y}$ can be found in the work of Press [3] and of King and Lasota [2]. Briefly, it can
be stated as follows : the integral expression for $G^{y}$ is invariant under each of the transformations $a \rightarrow (-a)$,
$B \rightarrow (-B)$, namely under inversions of the spin direction and the magnetic field direction. However, since the
sign of $G^{y}$ should flip under the combined transformations, $G^{y}$ must be identically zero. With this preparation, we now
can produce some numbers to censor the validity of this magnetic alignment process. Earlier, we argued that as mutually
complementary mechanisms, the magnetic alignment process precedes and the Blandford-Znajek process follows or else they may be
concurrent. To support this argument, it should be demonstrated that the time scale over which the alignment takes place is indeed
less than or comparable to that for the Blandford-Znajek process. To this end, we first note that
$(c^{5}/G^{2})(10^{15} G)^{-2}M_{\odot}^{-1} = 2.7\times 10^{3} (sec)$ has the right dimension for time scale. (Throughout this
work, we put, for the magnetic field strength, the number as big as $10^{15} (G)$ in order to make contact with the typical
observed luminosity of GRB in an attempt to explain it in terms of the magnetic alignment and/or Blandford-Znajek process.)
Thus recovering $G$ and $c$, the time scale for the magnetic alignment process turns out to be 
\begin{eqnarray}
\tau_{MA} = {3\over 2}B^{-2}M^{-1}(c^5/G^2) = {3\over 2}(2.7\times 10^{3})\left[B/10^{15}G\right]^{-2}\left[M/M_{\odot}\right]^{-1}
(sec).
\end{eqnarray}
Indeed, this is almost comparable to the time scale for the Blandford-Znajek process discussed in the literature [9]
\begin{eqnarray}
\tau_{BZ} &\sim & Mc^{2}/B^{2}r_{H}^{2}c \sim Mc^{5}/B^{2}M^{2}G^{2} \\
&=& c^{5}/G^{2}B^{2}M = (2.7\times 10^{3})\left[B/10^{15}G\right]^{-2}\left[M/M_{\odot}\right]^{-1} (sec) \nonumber
\end{eqnarray}
where the Kerr hole's horizon radius has been taken to be $r_{H} \sim GM/c^{2}$. Particularly note that for a solar-mass hole,
$M \sim M_{\odot}$ surrounded by an ultra strong magnetic field of order, say, $B \sim 10^{15} G$, the typical time scale for both
the Blandford-Znajek and the magnetic alignment processes is $\tau \sim 2.7\times 10^{3} (sec)$ which is to be compared to 
the duration of recently observed GRB spectrum [10]. We shall come back to this issue shortly. Next, we carry out the 
order-of-magnitude estimate for the power released during this magnetic alignment process, namely, the extracted rotational energy
per unit time due to the action of torque component $G^{z}$ in eq.(2). Certainly, in order for this new mechanism to attain a status 
of a viable model for the rotational energy extraction mechanism from a Kerr hole hopefully to be able to explain some observed spectrum
like that of GRB, it should pass the correct number generation test like this one. From our earlier review of the work by King and 
Lasota [2], the change in the magnitude of angular momentum of the hole over the alignment time scale is just the component of the 
torque along the hole's spin direction, $G^{z} = -(2/3)JB^{2}M \sin^{2}\gamma$. Thus it follows that :  \\
(1) the power loss at the horizon (resulting from the rotational energy lost by the hole) roughly is
\begin{eqnarray}
P_{hole} \sim |G^{z}|\Omega_{H} = {2\over 3}JB^{2}M\Omega_{H}\sin^{2}\gamma ,
\end{eqnarray}
(2) the power gained by the magnetic field would be
\begin{eqnarray}
P_{F} \sim |G^{z}|\Omega_{F} = {2\over 3}JB^{2}M\Omega_{F}\sin^{2}\gamma . 
\end{eqnarray}
(3) Lastly, the power converted into the hole's irreducible mass is then,
\begin{eqnarray}
P_{irr} = P_{hole} - P_{F} = |G^{z}|(\Omega_{H}-\Omega_{F}) = {2\over 3}JB^{2}M(\Omega_{H}-\Omega_{F})\sin^{2}\gamma 
\end{eqnarray}
where $\Omega_{H}$ and $\Omega_{F}$ are angular velocity of the hole and the magnetic field lines near the hole respectively.
Note here that the oblique magnetic field lines acquire the angular velocity $\Omega_{F}$ since, as we pointed out earlier,
the hole also exerts the torque $(-G^{z})$ on the magnetic field as a whole (which is equal and opposite to $G^{z}$ that the
magnetic field exerts on the hole) and as a result, this reaction torque spins up the swirl of the magnetic field lines.    
Finally, using the re-expressions of the quantities
\begin{eqnarray}
J &=& Ma = M^{2}\tilde{a} ~~~(0 < \tilde{a} \leq 1), \\
\Omega_{H} &=& {a\over (r^{2}_{+}+a^{2})} = {a\over 2Mr_{+}} \sim {M\tilde{a}\over 2M^{2}} \sim {\tilde{a}\over 2M} \nonumber
\end{eqnarray}
and taking the value of $\Omega_{F}$ as the ``optimal'' condition, $\Omega_{F}=\Omega_{H}/2$ to compare with the prediction of the 
Blandford-Znajek mechanism [9] in a moment, the power gained by the magnetic field lines which would eventually be released to infinite
distance can be estimated as
\begin{eqnarray}
P_{F} \sim {1\over 3}JB^{2}M\Omega_{H}\sin^{2} \gamma \sim ({1\over 6}\tilde{a}^{2}\sin^{2} \gamma)M^{2}B^{2}.
\end{eqnarray}
Thus noticing that $(G^{2}/c^{3})(10^{15}G)^{2}M^{2}_{\odot} = 6.7\times 10^{50} (erg/sec)$ has the right (Gaussian) dimension for
the power, we can rewrite the radiated power at the horizon in this magnetic alignment process as
\begin{eqnarray}
P_{MA} &\sim& ({1\over 6}\tilde{a}^{2}\sin^{2} \gamma)M^{2}B^{2}\left(G^{2}/c^{3}\right) \\
&=& 1.1\times 10^{50} \kappa_{MA}\tilde{a}^{2} \left[B/10^{15}G\right]^{2}\left[M/M_{\odot}\right]^{2} (erg/sec) \nonumber
\end{eqnarray}
with $\kappa_{MA} \equiv \sin^{2} \gamma \sim O(1)$ and $0 < \tilde{a} \leq 1$. It is remarkable that this result is indeed
comparable to the total radiated power of the Blandford-Znajek mechanism estimated in the literature [9]
\begin{eqnarray}
P_{BZ} = 1.7\times 10^{50} \kappa_{BZ}\tilde{a}^{2} \left[<B_{H}>/10^{15}G\right]^{2}\left[M/M_{\odot}\right]^{2} (erg/sec)
\end{eqnarray}
with $\kappa_{BZ} = f(h) \equiv \left[(1+h^{2})/h^{2}\right]\left[(h+1/h)\arctan (h)-1\right]$ $(0 < h \leq 1)$
which is again $\kappa_{BZ} \sim O(1)$ and $B_{H}=B_{\hat{r}}$ represents the magnetic field component normal to the horizon
as observed by a local observer, say, like the zero-angular-momentum-observer (ZAMO) [6]. Therefore, provided that strong
enough surface magnetic field of order, $B \sim 10^{15} (G)$ is available, even with the solar mass hole, $M\sim M_{\odot}$,
both the magnetic alignment and the Blandford-Znajek processes can generate the outgoing power of order 
$P \sim 10^{50} \kappa \tilde{a}^2 (erg/sec)$ with the typical time scale $\tau \sim 2.7\times 10^3 (sec)$ which is in reasonable
accord with some observed GRB spectrum [10], for example, GRB 990123 for which $E_{\gamma} = 3.4\times 10^{54} (\Omega_{\gamma}/4\pi)
(erg)$ or GRB 971214 for which $E_{\gamma} = 10^{53.5} (\Omega_{\gamma}/4\pi) (erg)$ in the time interval of order $\Delta \tau \sim
10^2 (sec)$, and we find the agreement quite impressive. Indeed, this agreement seems to have its basis on the condition that ultra
strong magnetic field, say, of order $B\sim 10^{15} (G)$ is actually available in the vicinity of the rotating hole. One can
imagine that this could be possible if the hole came out as a result of the gravitational collapse of a magnetar or else if it
accompanied nearby a young neutron star with large surface magnetic field. Discussions along this line can be found in [9].

\begin{center}
{\rm\bf IV. The magnetic alignment process - Microscopic version}
\end{center}

Thus far, we have discussed the estimate for the power released (particularly
at the horizon of the Kerr hole) during this magnetic alignment process by employing the expression for the torque exerted by an
asymptotically uniform magnetic field on a rotating hole given by King and Lasota [2]. This type of rotational energy extraction
process can be thought of as a ``macroscopic'' mechanism in that it is based upon an interaction of a rotating hole with a vacuum
(i.e., no plasma) electromagnetic field via a macroscopic torque. In association with this, it is interesting to note that several
years ago, there has been the study of rotational energy extraction mechanism of another type which is ``microscopic'' in nature
but still is based on the interaction between a rotating hole and a slowly-rotating vacuum electromagnetic field. Namely, some
time ago, Znajek [4] considered an environment in which a rotating hole is placed in a vacuum (i.e., without assuming the 
establishment of the force-free magnetosphere) electromagnetic field which is varying like $\sim \exp{(im\phi - i\omega t)}$.
Then the angular velocity of the field pattern around the hole is $\omega/m$. Indeed, one can imagine the possible occurrence of
such an environment.  For example [4], the gravitational collapse of a star that has formed a black hole could leave behind a cloud of
plasma orbiting round the hole carrying any large-scale non-axisymmetric magnetic field embedded in it. Alternatively, such a field
could be generated by a neutron star near the hole. And there is no reason why accretion should necessarily take place and hence it
may be that there are no significant charges and currents in the vicinity of the hole. Although this process considered by Znajek [4]
is totally different in nature from the ``superradiance'' discussed by Press and Teukolsky [11] appeared earlier than that, there is
one common implication and that is, provided $0< \omega/m < \Omega_{H}$, the hole will lose some of its (rotational) energy that will
eventually be radiated away to infinity. Essentially, Znajek was able to derive generally valid expressions for the rate of loss of
energy and angular momentum by the rotating hole to the field in terms of the toroidal fields at the event horizon. Indeed the way 
how he derived these quantities is similar to that in the context of the Blandford-Znajek mechanism. Namely, he assumed that the 
angular momentum and the rotational energy transport is achieved in terms of the conservation of {\it electromagnetic} angular
momentum and energy flux flowing from the hole to the nearby magnetic field but in this time not necessarily in the presence of the force-free
magnetosphere. The result is [4,5]
\begin{eqnarray}
|{dM\over dt}| = P_{z} = {2\over 3}JB^{2}_{\bot}M\Omega_{F}, 
~~~|{dL\over dt}| = {2\over 3}JB^{2}_{\bot}M
\end{eqnarray}
where $J=Ma$, $\Omega_{F}=\omega/m$ and $B_{\bot} = B\sin \gamma$ (with $\gamma$ being the angle between the asymptotic direction of 
the magnetic field and the 
hole's spin direction introduced earlier in eq.(2)) is the component of a generally oblique magnetic field perpendicular to the hole's
spin direction. Notice that this result of Znajek precisely agrees with our estimate for the radiated power at the horizon in the magnetic
alignment process given in eq.(6). And this result may imply that the type of rotational energy extraction mechanism considered by
Znajek can be thought of as a microscopic version of our magnetic alignment process presented in this work. 

\begin{center}
{\rm\bf V. Discussions}
\end{center}

To summarize, the magnetic
alignment process proposed in the present work (or its microscopic version studied by Znajek) may act as a complementary process to the
one like the Blandford-Znajek mechanism for the extraction of rotational energy from a Kerr hole and together, they can serve as a viable
model for the central engine of quasar, AGN, and even the GRB. Of course, in their spectra we actually observe, the contributions from
the two mechanisms, which we have found to be almost comparable, would be completely mixed and hence may well be indistinguishable which
part comes from which. Thus far, we have assumed that the magnetic field is uniform and constant
even at great distances from the hole (this is true for the microscopic process considered by Znajek as well). We now discuss what happens
if we relax this restriction and consider more realistic, asymptotically non-uniform field case. And to this end, we first recall the
celebrated Hawking's theorem [7] which can be stated concisely as follows. ``A stationary black hole must be either static (Schwarzschild) or
axisymmetric (Kerr).'' Thus when we apply this theorem to the case of Kerr hole immersed in a non-axisymmetric perturbing field, it can
be deduced that the hole must evolve in time, until either (i) it has lost all of its angular momentum and become a Schwarzschild hole,
or (ii) it has achieved an axisymmetric orientation with respect to the perturbing field, if exists. Thus coming back to our issue,
if the magnetic field is not uniform (not axisymmetric) far from the hole, then there is no possible final axisymmetric configuration to
be achieved and hence by the Hawking's theorem above, the hole must eventually lose all of its angular momentum or equivalently rotational
energy by transferring it to the magnetic field which will eventually be released as the form of, say, the charged particle emission
presumably along the magnetic field lines. The question here might be, then, again, between the magnetic alignment process and that of
Blandford-Znajek, which one will be actually at work even in this case with asymptotically non-uniform (not axisymmetric) magnetic
field. We speculate that it is likely that the both would be. One then might be puzzled how the magnetic alignment process could work
in this case when the magnetic field is asymptotically non-uniform. In his original work, Press [3] argued that the asymptotic
nonuniformity of the field would give rise to just higher-order effect which renders parallel angular momentum component of the hole
as well as the perpendicular one dissipate, namely that a mechanism which is almost basically equivalent to the magnetic alignment process
would still operate. And it is this point that allows us to believe that the magnetic alignment process would have essentially the same
chance to work as the Blandford-Znajek process would even under a random and hence more realistic circumstances like this. To conclude, the
magnetic alignment process proposed in this work and the Blandford-Znajek mechanism appear to be on nearly equal footing in serving as basic
mechanisms for the rotational energy extraction from a rotating black hole. Moreover, as we suggested earlier, these two processes together
can collaborate and serve as a viable model for the central engine of GRB. In this regard, there is one more issue to be clarified and that is,
particularly for the magnetic alignment process, what exactly the working mechanism for the subsequent release of the extracted rotational
energy to infinitely far region would be. Earlier, we speculated that it still would be via the charged particle (plasma) emission
presumably along the magnetic field lines much like in the context of the Blandford-Znajek process. This type of argument, however, may seem
inconsistent since for the environment set by the magnetic alignment process (or its microscopic version studied by Znajek), the rotating
hole is surrounded by a vacuum (i.e., no plasma) electromagnetic field. If, however, the magnetic field is strong enough in order for the
Blandford-Znajek or the magnetic alignment process to be able to account for the observed energy spectrum of GRB as we remarked earlier,
this naive picture needs to be modified. Namely, as Blandford and Znajek [1] remarked, if the field strength is large enough, say, of order
$B\sim 10^{15} (G)$, the vacuum surrounding the hole becomes unstable because any stray charged particles will be electrostatically
accelerated and will radiate and this radiation will in turn produce further charged particles in the form of electron-positron pairs.
When charges are produced so freely, the electromagnetic field around the horizon will become approximately force-free, namely a force-free
magnetosphere will be established. Therefore in reality, if the surrounding magnetic field is strong enough, charged particles are available
due to the cascade production of electron-positron pairs and hence the extracted rotational energy in the magnetic alignment process discussed
in the present work may be released in the form of charged particle emission as well. Lastly, it is also interesting to note that since the
magnetic alignment process proposed in the present work is {\it time dependent} by its nature in contrast to the Blandford-Znajek process,
it may well generate gravitational waves which would carry off the rotational energy of the black hole. Therefore, the generation of the
gravitational waves involved in this magnetic alignment process can be considered as a possible precursor for GRB. This possibility appears
to deserve serious attention and is being examined by us.
\\
\\
H. Kim was financially supported by the BK21 Project of the Korean Government and H.K.Lee and C.H.Lee
were supported in part by grant No. R01-1999-00020 from the Korea Science and Engineering Foundation.

\vspace*{2cm}

\noindent

\begin{center}
{\rm\bf References}
\end{center}

\begin{description}

\item {[1]} R. D. Blandford and R. L. Znajek, Mon. Not. R. Astron. Soc. {\bf 179}, 433 (1977) ; For recent study of its theoretical aspects,
            see, H. Kim, Chul H. Lee, and H. K. Lee, Phys. Rev. {\bf D63}, 064037 (2001) ; H. Kim, H. K. Lee, and Chul H. Lee, {\it ibid.}
            {\bf D63}, 104024 (2001).
\item {[2]} A. R. King and J. P. Lasota, Astr. Astrophys. {\bf 58}, 175 (1977).
\item {[3]} W. H. Press, Astrophys. J. {\bf 175}, 243 (1972). 
\item {[4]} R. L. Znajek, Mon. Not. R. Astron. Soc. {\bf 179}, 457 (1977).
\item {[5]} M. D. Pollock, Proc. Roy. Soc. Lond. {\bf A350}, 239 (1976).
\item {[6]} J. M. Bardeen,  Astrophys. J. {\bf 162}, 71 (1970) ; J. M. Bardeen, W. H. Press, and S. A. Teukolsky, {\it ibid.} {\bf 178}, 347
            (1972).
\item {[7]} S. W. Hawking, Comm. Math. Phys. {\bf 25}, 152 (1972).
\item {[8]} C. W. Misner, K. S. Thorne and J. A. Wheeler, {\it Gravitation} (San Francisco : Freeman, 1973).
\item {[9]} H. K. Lee, R. A. M. J. Wijers, and G. E. Brown, Phys. Rep. {\bf 325}, 83 (2000).
\item {[10]} S. R. Kulkarini et al., Nature (London) {\bf 393}, 35 (1998), astro-ph/9902272. 
\item {[11]} W. H. Press and S. A. Teukolsky,  Nature (London) {\bf 238}, 211 (1972).
\item {[12]} P. Goldreich, and W. H. Julian, Astrophys. J. {\bf 157}, 869 (1969). 
\item {[13]} K. S. Thorne, R. H. Price, and D. A. Macdonald, {\it Black Holes: The Membrane Paradigm} (Yale University
             Press, New Haven and London, 1986).
\item {[14]} J. M. Bardeen, and J. A. Petterson, Astrophys. J. {\bf 195}, L65 (1975).

\end{description}

\end{document}